# Data-driven modelling of low-dimensional dynamical structures underlying complex full-body human movement

Ryota Takamido[1]*, Chiharu Suzuki[1], Hiroki Nakamoto[2]

1. Sports Innovation Organization, National Institute of Fitness and Sports in Kanoya, Kanoya, Kagoshima 891-2393, Japan

2. Faculty of Physical Education, National Institute of Fitness and Sports in Kanoya, Kanoya, Kagoshima 891-2393, Japan

Corresponding author:

Email: rtakamido@nifs-k.ac.jp (RT)

**Abstract**

One of the central challenges in the study of human motor control and learning is the degrees-of-freedom problem. Although the dynamical systems approach (DSA) has provided valuable insights into addressing this issue, its application has largely been confined to cyclic or simplified motor movements. To overcome this limitation, the present study employs neural ordinary differential equations (NODEs) to model the time evolution of non-cyclic full-body movements as a low-dimensional latent dynamical system. Given the temporal complexity full-body kinematic chains, baseball pitching was selected as a representative target movement to examine whether DSA could be extended to more complex, ecologically valid human movements. Results of the verification experiment demonstrated that the time evolution of a complex pitching motion could be accurately predicted ($R^2 > 0.45$) using the NODE-based dynamical model. Notably, approximately 50% of the variance in the latter half of the pitching motion was explained using only the initial ~8% of the temporal sequence, underscoring how subsequent movement evolves from initial conditions according to ODE-defined dynamics in latent space. These findings indicate the potential to extend the DSA to more complex and ecologically valid forms of human movement.

**Keywords:** Dynamical system, Degree-of-freedom problem, Human movement, Machine learning, Neural ordinary differential equations

# 1. Introduction

One of the central challenges in the study of human motor control and learning is the degrees-of-freedom (DoF) problem [1]. Because the human body is a highly redundant motor system comprising hundreds of joints, many combinations of joint angles can achieve a given movement goal.

The dynamical systems approach (DSA) conceptualizes complex human movement as the time evolution of a low-dimensional dynamical system regulated by underlying self-organizing processes [2-4]. In this framework, collective variables summarize the coordination among numerous degrees of freedom, and their temporal evolution is described by nonlinear differential equations. This approach has successfully explained stability, phase transitions, and coordination phenomena in cyclic movements such as bimanual coordination [5]. By focusing on emergent coordination dynamics [6], DSA provides insights that purely computational or optimization-based frameworks cannot naturally capture[7].

Nonetheless, a limitation of current DSA studies is the narrow range of target movements examined. Specifically, prior research has primarily analyzed coordination dynamics in cyclic movements, such as bimanual coordination and gait [5, 8], or simple discrete movements, including reaching [9-10]. However, real-world human movement encompasses a broader spectrum of high-DoF motor behaviours. Although DSA has been applied to more complex movements, such as full-body athletic movements in sports contexts [11-14], these studies have largely focused on coordination dynamics among a limited set of variables (e.g., the head or center of mass), rather than deriving a mathematical representation of coordination across complex sequences of full-body joint movements. Importantly, this limited application may reflect multiple factors, including challenges in designing appropriate differential equation structures, and the fact that many conventional analytical methods were developed under assumptions of periodicity or stationarity.

Several studies have explored extending DSA to more complex movements, such as full-body athletic tasks. First, even highly complex full-body movements often exhibit stereotyped coordination patterns that can be effectively compressed into low-dimensional representations using dimension-reduction techniques, such as principal component analysis (PCA) [15, 16]. Furthermore, in sequential full-body movements, inertia is critical, such that motor actions at a given phase depend strongly on the preceding phases [17, 18]. Thus, the temporal evolution of these sequential movements may be predictable using differential equations that incorporate such temporal dependencies. Moreover, despite their biochemical complexity and kinematic redundancy, full-body movements often exhibit high reproducibility and robustness to perturbations [19]. This suggests that underlying stable coordination dynamics can be described within a dynamical systems framework.

Based on these considerations, the limited scope of DSA applications to date is likely attributable not to conceptual limitations of the framework itself but rather to technical challenges in the mathematical modeling of complex, noncyclic, full-body movements. In this context, Neural Ordinary Differential Equations (NODEs) [20] may offer an effective solution to the technical limitations of current DSA approaches. NODEs are a neural-network-based mathematical modeling technique that represent the time evolution of features within a latent space, typically constructed using methods such as variational autoencoders (VAEs) [21], via a neural-network-based formulation of differential equations. A key advantage of this approach is its ability to capture the temporal structure of complex human movements in a low-dimensional latent feature space, without relying on periodicity or stationarity assumptions. Previous research has demonstrated the effectiveness of NODEs across a range of human motion prediction tasks, consistently reporting high performance [22, 23]. Unlike traditional DSA implementations that require predefined coordination variables, NODEs learn governing differential equations directly from data, enabling the modelling of non-cyclic, full body movements as the evolution of low-dimensional dynamical systems in latent space.

The present study aimed to investigate whether coordination dynamics in non-cyclic, full-body movements can be modelled and predicted using NODEs. Baseball pitching was selected as a representative task. As demonstrated in previous studies [24], one of the most prominent characteristics of baseball pitching is a complex kinematic chain that progresses sequentially from the lower limbs to the trunk and upper limbs. If this complex movement can be effectively captured as latent dynamics within a low-dimensional space, it would suggest that DSA is extensible beyond its conventional application domains. The source code used in this study, the anonymized dataset, and other supporting materials are publicly available on GitHub (https://github.com/takamido/NODE_baseball_pitching).

## 2. Methods

### 2. 1. NODE model used in this study

### 2.1.1 Overview

Figure 1 illustrates an overview of the NODE model used in this study. This model aims to predict the temporal evolution of target movement using information from its initial state as input. If complex full-body movements can be accurately modeled as the temporal evolution of a dynamical system in a low-dimensional latent space, their future evolution is predictable with high precision based solely on the initial state.

During model prediction, the original motion data were first input into the encoder (Figure 1a). From this sequence, only the initial portion of the motion (e.g., the first 8% of the time series) was used to determine the initial position in the low-dimensional latent space (Figure 1b). We then generated latent trajectories by integrating ODE learned by the neural network (Figure 1c). Finally, the generated trajectory was passed to the decoder, which reconstructed the full-body motion based on the predicted latent trajectory (Figure 1d). The differential-equation structure was learned in a data-driven manner, eliminating the need for predefined collective variables. By parameterizing dynamics with a neural network, the model captured low-dimensional movement structure without assuming periodicity or stationarity, thereby overcoming key limitations of traditional DSA. The following sections provide the detailed theoretical and technical descriptions of each component of the proposed model.

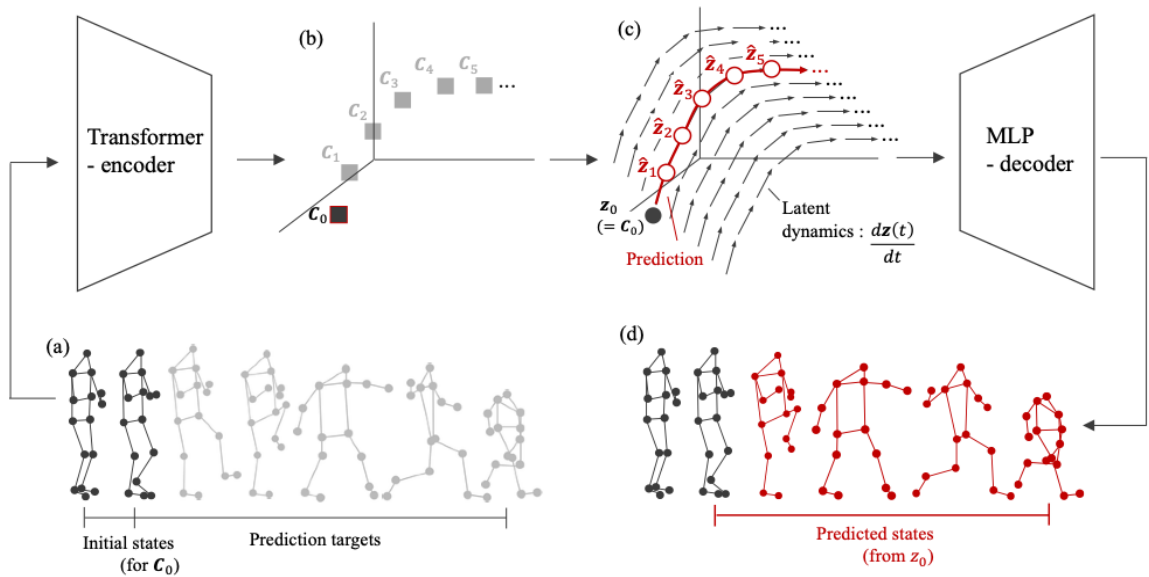

Figure 1. Overview of the NODE model used in this study for predicting the pitching movement.

**2.1.2. Transformer encoder for latent space generation**

Here, a transformer module [25] was selected as the encoder to embed human motion data into a low-dimensional latent space. This choice was motivated by findings from previous research [26], which suggested that the transformer module is effective for encoding and decoding temporal dependencies characteristic of full-body, long sequential movements. Because the present study primarily focuses on applying the transformer architecture to motion generation, a detailed description of its internal components is omitted. Additional details are provided in the original implementation [25].

The encoding module receives motion data and maps it to a lower-dimensional latent space, producing a latent trajectory represented as a sequence of latent states. Theoretically, let input feature $x_{0:T-1} \in \mathbb{R}^{T \times D}$ denote a time series of the whole-body human motion data of length $T$, where $D$ is the dimension of motion features per frame (e.g., concatenated 3-dimensional joint positions). The encoder maps the $x_{0:T-1}$ into a sequence of $K$ low-dimensional latent variables $z_{0:K-1} \in \mathbb{R}^{K \times d}$. Each latent variable captured the movement state at specific time point. Specifically, full-body pitching movements were mapped onto 12 trajectory points in a three-dimensional latent space. Among these variables, the initial state $z_0$ is used to generate the latent trajectory via the ODE (Figure 1b and c). During latent-variable generation, a causal mask is applied to prevent access to future information.

**2.1.3 ODE for modelling the time evolution of latent states**

From the initial latent state $z_0$ obtained from the encoder, the subsequent evolution of the system in the latent space is described as a deterministic flow defined using the following ODE:

$$\frac{d\hat{z}(t)}{dt} = f_\theta(z(t), t), \tag{1}$$

where $f_\theta$ is a neural network (e.g., multilayer perceptron [MLP]) that defines a vector field on the latent space. As illustrated in Figure 1c, future states are predicted deterministically from an initial condition $z_0$ by following the flow induced by Eq. (1). This deterministic ODE flow captures the high reproducibility and stereotyped structure underlying complex full-body movements.

In the proposed model, the ODE predicts a latent trajectory with the same length as the original time series ($T$) from the initial latent state $z_0$, yielding $\hat{z}_{1:T-1} \in \mathbb{R}^{T \times d}$. The predicted trajectory is then passed through a decoder to reconstruct the corresponding sequence of motion states in the original feature space. Finally, an MLP decoder reconstructs the original motion sequence (Figure 1d). Because the encoder compresses full-body motion into a low-dimensional representation, a relatively simple MLP suffices for decoding.

### 2.1.4 Design of the loss function

Following the standard VAE formulation, the model was trained end-to-end by minimizing the following loss function:

$$\mathcal{L} = \lambda_{recon}\mathcal{L}_{recon} + \lambda_{kl}\mathcal{L}_{kl}, \quad (2)$$

where $\mathcal{L}_{recon}$ denotes the reconstruction loss weighted by the hyperparameter $\lambda_{recon}$ for the original motion sequence. $\mathcal{L}_{kl}$ is the Kullback-Leibler divergence loss weighted by the $\lambda_{kl}$ that regularizes the latent-variable posterior toward a standard normal prior [27]. By penalizing the reconstruction error and the posterior distribution shape, the latent space is encouraged to exhibit a smooth and continuous structure, which facilitates the definition of a well-behaved dynamical flow in the latent space.

### 2.2 Dataset

The applicability of the NODE model described in the previous section was evaluated using baseball pitching data as a representative noncyclic, full-body movement. The dataset comprised measurements from eight collegiate-league pitchers (174.1 ± 4.1 cm, 74.91 ± 3.8 kg, 19.87 ± 1.0 years). All pitchers had more than 10 years of playing experience. Previously collected data were used in this study. An opt-out informed consent framework was adopted in accordance with the university's ethical guidelines. Information regarding the study, including its purpose and usage, was made publicly available to enhance transparency. Individuals whose data were included were provided a clear opportunity to decline participation.

For dataset collection, each pitcher's motion was recorded at a sampling rate of 200 Hz using 16 synchronized optical motion-capture cameras (Raptor-E and Kestrel 2200, Motion Analysis Corp., Santa Rosa, USA). To simulate realistic game-like conditions, pitchers threw diverse pitch types to random target locations. All pitches were delivered using a wind-up motion characterized by a large leg lift. From the captured data, the positions of 15 joints were extracted as follows: the parietal region (head), bilateral acromions (shoulders), lateral epicondyles of the humerus (elbows), radial styloid processes (wrists), greater trochanters of the femur (hips), lateral femoral condyles (knees), and both heels and toes (tops of the shoes).

Based on the measured joint position data, the analysis window for each pitch was determined using the following procedure. For each pitch, the time at which the pitcher's leading knee reached its maximum height was identified. Subsequently, by tracing backward in time from this point, the earliest frame at which the knee's vertical upward velocity dropped below 5% of its maximum value was selected as the onset of motion (i.e., initial time frame). Next, the time point of maximum right-wrist velocity was detected and defined as ball release. Using the trial in which ball release occurred latest as a temporal reference, the motion sequences for all trials were temporally aligned and extracted accordingly. Consequently, time-series data representing the time evolution of the pitching motion, comprising 45-dimensional features (15 joints × 3D positions), were obtained.

In conventional biomechanical studies, the time between motion onset and ball release is typically normalized to a fixed scale (e.g., 0–100%) for analysis (e.g., [28]). Unlike many dynamical systems studies, in this study, we did not apply temporal normalization, preserving the natural temporal structure of the movement. This allowed the latent dynamics to be modelled in real physical time, rather than in an artificially rescaled phase representation. Consequently, some trials included a follow-through phase beyond ball release, whereas others terminated precisely at ball release. This results in a substantial increase in inter-trial variability during the latter portion of the motion sequence, which makes the prediction task more challenging. Therefore, this dataset provides a meaningful opportunity to investigate whether NODEs can be used to model trial-to-trial variability in individual-specific dynamical structure. For reference, a video that overlays all pitching motions from all pitchers has been uploaded to the GitHub repository as supporting information. Through the above data generation process, eight individual datasets were constructed, one per pitcher, as summarized in Table 1.

Table 1. Participants' information for the analysis

| Participant ID | Pitch number | Time frame |
|---|---|---|
| sub01 | 105 | 450 |
| sub02 | 100 | 396 |
| sub03 | 81 | 488 |
| sub04 | 137 | 513 |
| sub05 | 140 | 447 |
| sub06 | 115 | 503 |
| sub07 | 140 | 359 |
| sub08 | 150 | 391 |

## 2.3 Experiment

Based on the developed dataset, a NODE model was trained to capture the dynamics of the pitching motion, and its ability to predict temporal evolution was evaluated. During training, the transformer encoder architecture used transformer modules with three layers, eight attention heads per layer, and a model dimension of 256. The latent-space dimensionality was set to $d = 3$, thereby reducing the original motion data with dimension $D = 45$ to a lower-dimensional representation. The entire motion sequence was modeled using three latent features. The number of time-series points mapped onto the latent space by the transformer encoder was set as $K = 12$. Consequently, the initial latent state $z_0$ encodes approximately the first 8% of the original motion sequence and serves as the initial condition of the dynamical system governing the subsequent evolution. In the prediction task, the model forecasts the remaining 92% of the motion using only the initial information. The latent dynamics $f_\theta$ (i.e., the vector field) was modeled using a three-layer MLP with two hidden layers, each consisting of 128 units and tanh activation functions. The MLP decoder comprised three fully connected layers, with two hidden layers of 256 units and ReLU activations, followed by a linear output layer producing 45-dimensional joint positions.

Training was performed using 10-fold cross-validation for each pitcher dataset. In each fold, nine subsets were used for training and one subset was used for testing. Motion features were standardized using the mean and standard deviation calculated from the training set. The weights for each penalty term in the loss function were set as $\lambda_{kl} =$

$1 \times 10^{-3}$ and $\lambda_{recon} = 1.0$. The batch size was set to 32. With these hyperparameter settings, the model was trained using the Adam optimizer with a learning rate of $1 \times 10^{-4}$. The model was trained for 1500 epochs. Training was conducted on the Google Colab platform using an NVIDIA T4 GPU.

To evaluate the prediction performance of the trained model, the following metrics were computed. First, reconstruction errors for the test motions were quantified using the root mean squared error (RMSE) at each predicted time frame and were aggregated for each joint across all pitchers and cross-validation folds. To ensure reproducibility, the latent variables were deterministically derived at test time using the mean of the encoded latent distribution. For comparison, RMSE was computed for a baseline model that predicted each frame using the mean of the training data. To evaluate whether the learned dynamical flow constrains the evolution of the movement beyond statistical baselines, the time-resolved coefficient of determination, $R^2(t)$ was computed using the following equation:

$$R^2(t) = 1 - \frac{SS_{res}(t)}{SS_{ori}(t)}, \tag{3}$$

where $SS_{ori}(t)$ is the total sum of squares of the original data at each time frame, and $SS_{res}(t)$ is the residual sum of squares between the original data and the model predictions. To evaluate prediction accuracy, particularly in the latter phase of motion where variability was often greater, both the mean $R^2$ value across the entire time series and the mean $R^2$ value across the latter 50% were calculated for each individual.

## 3. Results

Figure 2 displays the RMSE values at each time point for each participant. Baseline predictions based on the training-data mean exhibited a pronounced increase in error during the later phase of the motion (Figure 2). In contrast the NODE-based prediction mitigated this error growth, suggesting that the learned latent dynamics shape the later temporal evolution of the movement (Figure 2). Across all pitchers and the entire prediction interval, the mean RMSE was 66.1 ± 15.7 mm. Figure 3 presents the $R^2$ values for each participant at each timepoint. In Figure 3, the NODE-based prediction, conditioned only on approximately 8% of the initial motion, substantially reduced uncertainty in the subsequent time evolution of the full-body pitching movement, indicating that the learned latent dynamics constrained the trajectory beyond what would be expected from trial-averaged behaviour. The mean $R^2$ across all pitchers over the entire motion was 0.46 ± 0.20, and the mean $R^2$ computed for the second half of the motion was 0.49 ± 0.20. Figure 4 displays a comparison of the original and reconstructed motions for all pitching trials for each participant. The NODE model captures trial-to-trial variability in the subsequent evolution of the motion based on subtle differences in the preparatory phase, consistent with sensitivity to initial latent states (Figure 4). Finally, Figure 5 presents an example of the learned latent trajectories and the corresponding motions generated from the model. The temporal evolution of the pitching movement is largely captured by a three-dimensional latent dynamical system (Figure 5).

For reference, videos of the reconstructed motions for all available pitchers were created, along with aligned comparison videos of the learned latent trajectories, and provided as Supporting Information on the GitHub repository.

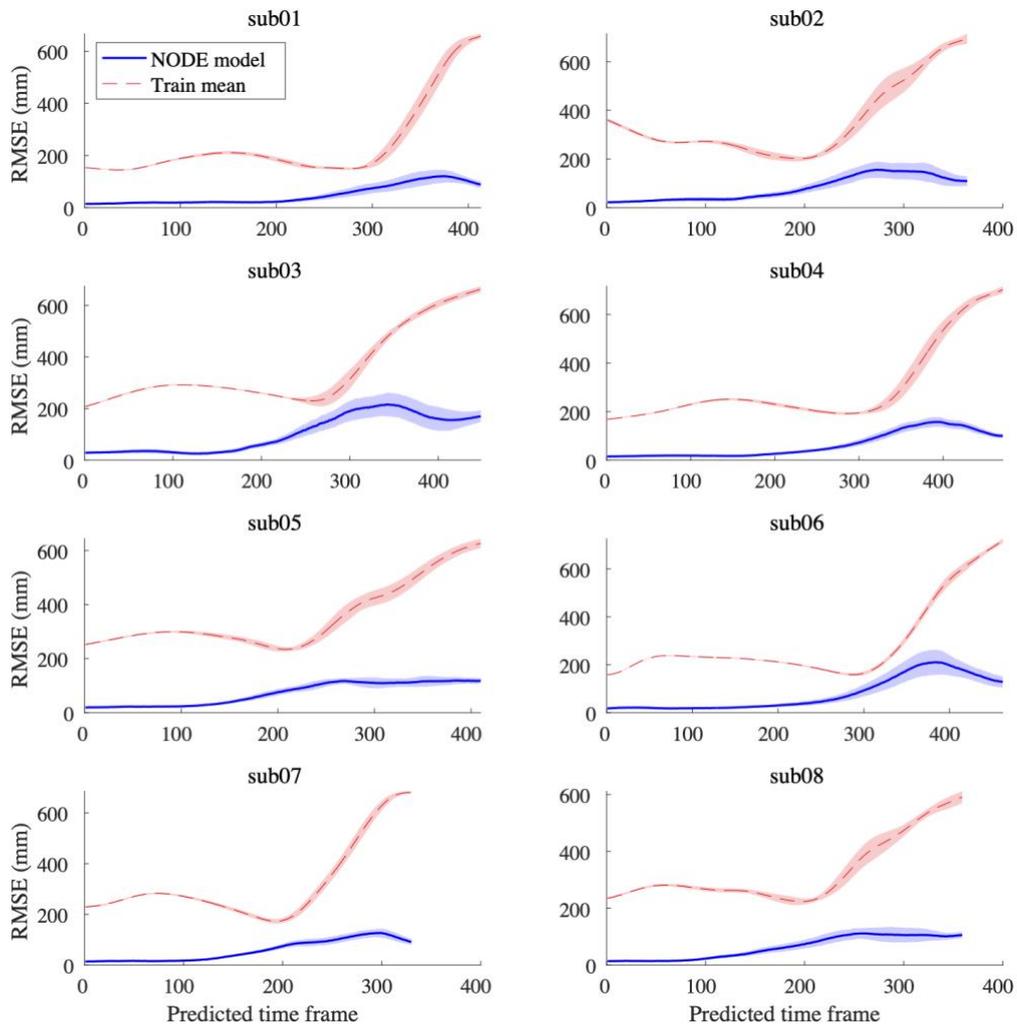

Figure 2. RMSE at each predicted time point for each participant. Mean ± standard deviation across the 10 cross-validation folds.

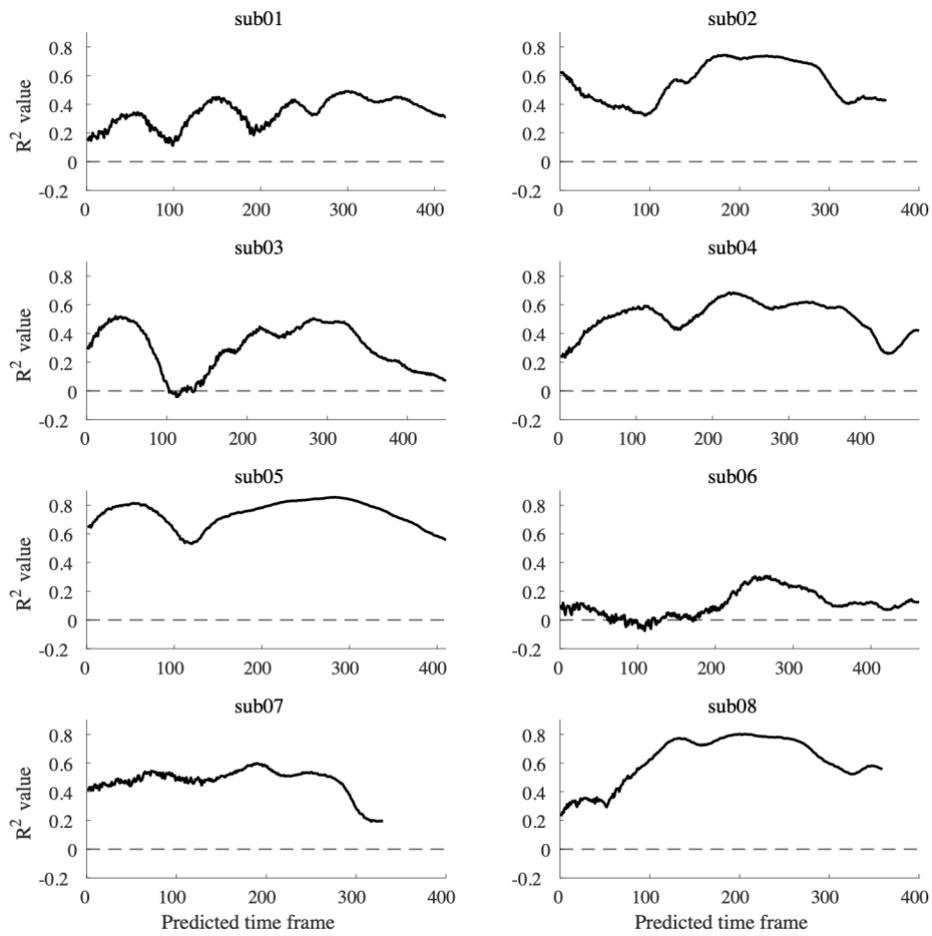

Figure 3. $R^2$ value at each predicted time point for each participant. Mean ± standard deviation across the 10 cross-validation folds.

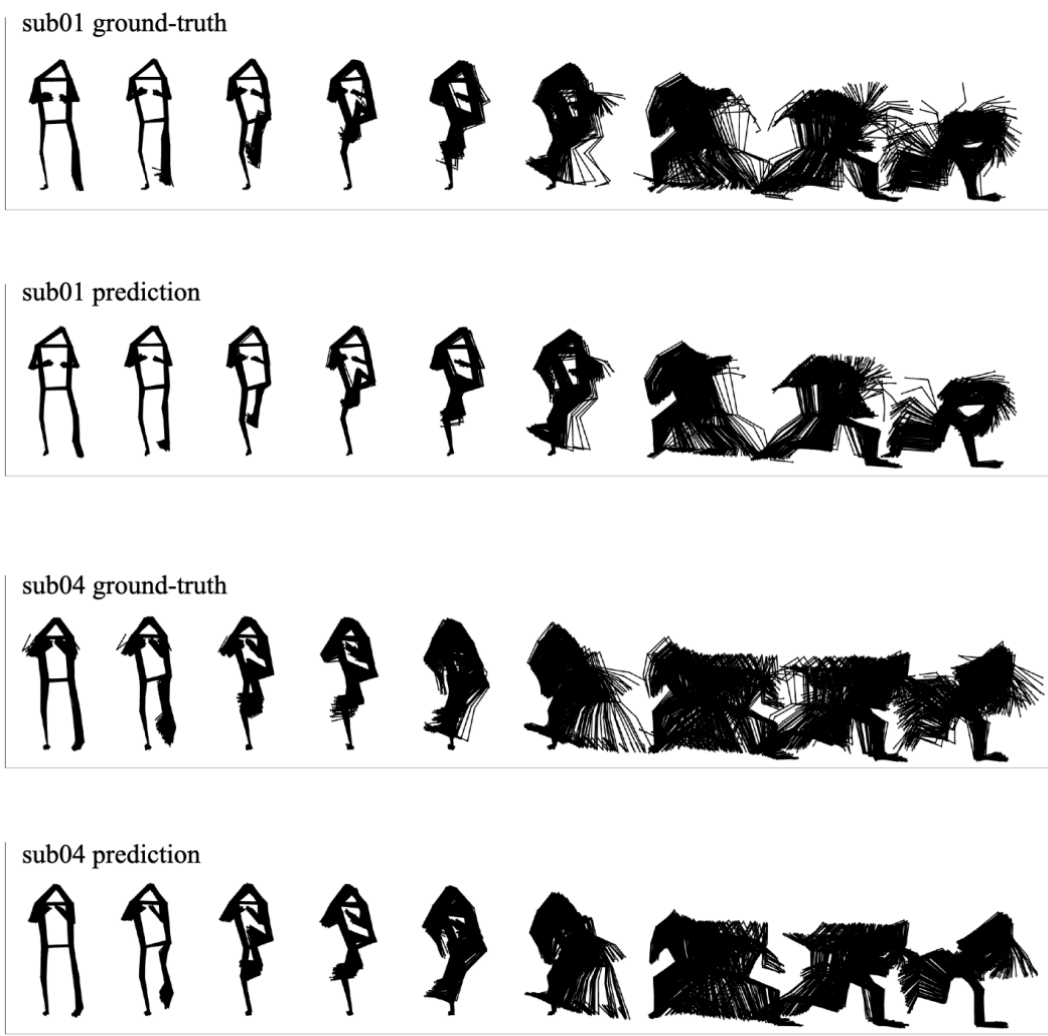

Figure 4. Comparison of motion patterns before and after reconstruction. Ground-truth and predicted trajectories are presented for all pitching trials. Data from sub01 and sub04 are displayed as representative examples of a pitcher with relatively stable motion patterns (sub01) and one with more variable patterns (sub04), respectively.

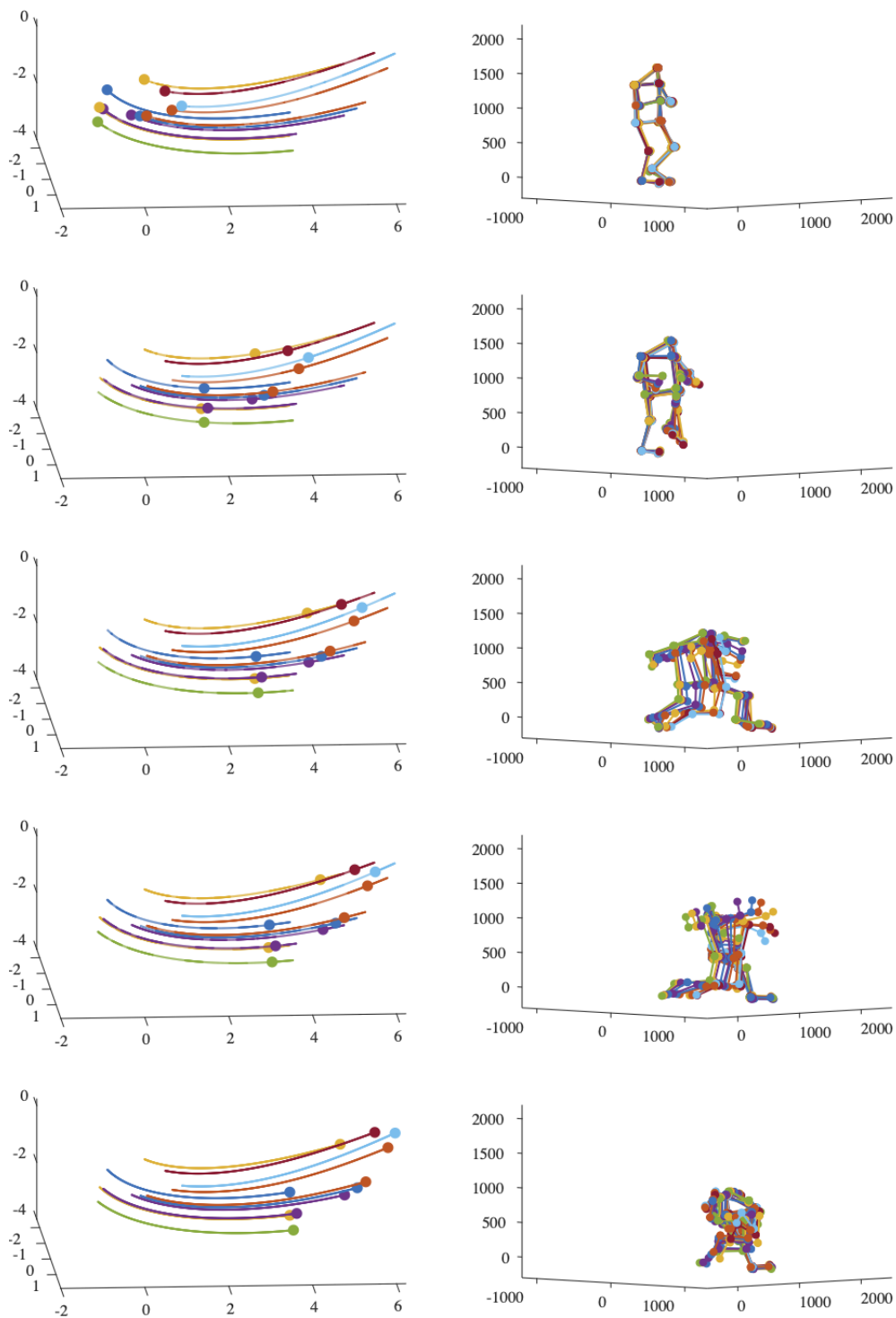

Figure 5. Relationship between learned latent trajectories (left column) and corresponding generated motions (right column). Data from Participant 1 in Fold 1 are presented as a representative example.

## 4. Discussion

The high predictive performance of the NODE model demonstrates that the evolution of full-body pitching movements follows a three-dimensional dynamical system in latent space. Although the pitching movement can seem complex, the dynamics of its time evolution can be effectively captured by a three-dimensional flow in the latent space. Therefore, modeling the pitching motion—a representative example of a noncyclic, high-DoF, full-body movement—using NODE in this study demonstrated the potential to expand the applicability of DSA.

It is worth considering why the NODE model achieved high prediction accuracy. First, movement analysis was performed by trained athletes, whose motor patterns are highly stereotyped and stable through deliberate practice [29]. Previous studies have suggested that repeated practice can reorganize coordination and, in some task contexts, reduce the effective dimensionality of movement, leading to the emergence of more stable low-dimensional dynamical structures [30]. Regarding DSA, such pattern formation enhances stability against perturbations [31]; thus, experienced pitchers, who must meet high demands for accuracy and reproducibility, may have adapted to these task constraints [32] by developing more stabilized dynamical structures. Accordingly, the model may have greater difficulty predicting movements in novice individuals, whose movement patterns tend to be less stable [33].

Nonetheless, some movement characteristics were not captured by the model developed in this study. The key limitation is movement variability reduction. The generated motions appeared more stereotyped, often translating similar postural patterns or shifting phases (Figure 4). Therefore, the remaining 50% of the motion variance may be attributed to fluctuations that are not incorporated into the present model. From a DSA perspective, fluctuations are not merely random noise but reflect the stability properties of the underlying coordination dynamics [34]. While the present deterministic latent dynamical model captures the dominant low-dimensional flow underlying the full-body movement, it does not explicitly model the fluctuation dynamics around this structure.

From a practical perspective, the tendency of the model to generate more stereotyped movements may be interpreted as a lack of functional variability in the generated motion. Functional variability refers to structured, task-relevant variations in movement that allow flexibility and adaptability while maintaining overall performance stability [35]. Because functional variability plays a key role in the adaptive motor control of high DoF human movement, representing such fluctuations in mathematical models would constitute a fundamental issue. Although stochastic fluctuations can be introduced into the generated movements through approaches such as extensions to stochastic differential equations [36], more sophisticated models are necessary to capture the functional covariance structures embedded in human movement.

It is important to clarify how the presented approach differs from standard dynamical systems modelling frameworks, such as the HKB model [5]. Standard DSA studies identify parameters (e.g., relative phase between both hands) and demonstrate their stability using systematic manipulation of control variables (e.g., oscillation frequency), thereby revealing phenomena such as phase transitions and metastability [34]. In contrast, the present NODE-based model infers a low-dimensional latent dynamical structure without explicitly defining order or control parameters. This demonstrates canonical DSA phenomena such as phase transitions or metastability, but in providing empirical evidence that is ecologically valid, high-DoF full-body movements admit a substantial low-dimensional dynamical structure. In this sense, the work may be viewed as a data-driven extension that complements, rather than replaces, standard order-parameter-based DSA modelling.

Finally, this study has some limitations and directions for future work. First, in accordance with the traditional DSA framework, it would be necessary to systematically and continuously manipulate control parameters, such as environmental constraints (e.g., pitching distance), and examine whether the latent variables exhibit phase transitions or hysteresis [34]. Furthermore, to better interpret the meaning of the identified latent variables, we must mathematically analyze learned latent space by the NODE [37] would be informative. Finally, from a practical perspective, comparing the NODE model prediction accuracy between experts and novices would be useful for understanding the roles of stereotypicality and functional variability in motor learning.

## 5. Conclusion

In conclusion, the high predictive performance of the NODE model suggests that complex, non-cyclic full-body movements can be described within a low-dimensional dynamical systems framework. Notably, approximately 50% of the variance in the latter half of the pitching motion was explained using only the initial ~8% of the movement sequence. This underscores the extent to movement evolves from its initial state according to the ODE-defined dynamical flow in latent space. These findings indicate the potential to extend the DSA to more complex and ecologically valid forms of human movement. Incorporating noise and structured functional variability in future models may enable more faithful reproduction of the rich dynamics inherent in complex full-body actions. Furthermore, systematic manipulation of control parameters and observation of stability changes in identified parameters would facilitate rigorous validation within the formal DSA framework.